\documentclass[graybox]{svmult}

\usepackage{mathptmx}       
\usepackage{helvet}         
\usepackage{courier}        
\usepackage{type1cm}        
\usepackage{makeidx}         
\usepackage{graphicx}        
\graphicspath{{Figures/}}
\usepackage{multicol}        
\usepackage[bottom]{footmisc}

\makeindex             

\graphicspath{{/}{fig/}}
\usepackage{multirow}
\usepackage{booktabs}
\usepackage{breqn}
\usepackage{tabularx}

\usepackage{caption}
\usepackage{subcaption}

\usepackage{float}
\usepackage{hyperref}

\usepackage{pgfplots}
\pgfplotsset{compat=1.7}
\usepgfplotslibrary{groupplots}

\usepackage{pifont}
\newcommand{\cmark}{\ding{51}}%
\newcommand{\xmark}{\ding{55}}%

\newlength\figureheight
\newlength\figurewidth
\begin{document}

\title*{Blockchain for Mobile Edge Computing: Consensus Mechanisms and Scalability}
\titlerunning{Blockchain for MEC: Consensus Mechanisms and Scalability}
\author{Jorge Pe\~{n}a Queralta and Tomi Westerlund}

\institute{Jorge Pe\~{n}a Queralta \at \href{https://tiers.utu.fi}{Turku Intelligent Embedded and Robotic Systems Lab, University of Turku, Turku, Finland} \\\email{jopequ@utu.fi}
\and Tomi Westerlund \at \href{https://tiers.utu.fi}{Turku Intelligent Embedded and Robotic Systems Lab, University of Turku, Turku, Finland} \\\email{tovewe@utu.fi}}
%
%
\maketitle


\abstract*{Mobile edge computing (MEC) and next-generation mobile networks are set to disrupt the way intelligent and autonomous systems are interconnected. This will have an effect on a wide range of domains, from the Internet of Things to autonomous mobile robots. The integration of such a variety of MEC services in a inherently distributed architecture requires a robust system for managing hardware resources, balancing the network load and securing the distributed applications. Blockchain technology has emerged a solution for managing MEC services, with consensus protocols and data integrity checks that enable transparent and efficient distributed decision-making. In addition to transparency, the benefits from a security point of view are evident. Nonetheless, blockchain technology faces significant challenges in terms of scalability. In this chapter, we review existing consensus protocols and scalability techniques in both well-established and next-generation blockchain architectures. From this, we evaluate the most suitable solutions for managing MEC services and discuss the benefits and drawbacks of the available alternatives.}

\abstract{Mobile edge computing (MEC) and next-generation mobile networks are set to disrupt the way intelligent and autonomous systems are interconnected. This will have an effect on a wide range of domains, from the Internet of Things to autonomous mobile robots. The integration of such a variety of MEC services in a inherently distributed architecture requires a robust system for managing hardware resources, balancing the network load and securing the distributed applications. Blockchain technology has emerged a solution for managing MEC services, with consensus protocols and data integrity checks that enable transparent and efficient distributed decision-making. In addition to transparency, the benefits from a security point of view are evident. Nonetheless, blockchain technology faces significant challenges in terms of scalability. In this chapter, we review existing consensus protocols and scalability techniques in both well-established and next-generation blockchain architectures. From this, we evaluate the most suitable solutions for managing MEC services and discuss the benefits and drawbacks of the available alternatives.}


\section{Introduction}

The scope of the Internet of Things (IoT) has been growing over the past decade, encompassing an ever larger ecosystem that spans multiple domains. Some of the most prominent research directions are smart cities~\cite{da2014internet, salman2018iot}, vehicular technology~\cite{wu2018spatial, queralta2019collaborative}, or smart healthcare systems~\cite{doukas2012bringing, mutlag2019enabling, queralta2019edgeai}. In all these domains, a common factor is that IoT systems are evolving towards more distributed architectures~\cite{edwards2016hajime}. This shift from more traditional cloud-centric architectures has crystallized in the edge computing paradigm~\cite{hu2015mobile, shi2016edge, qingqing2019fpga}. At the same time, novel technologies are increasingly designed with decentralization in mind from their inception. Among these, blockchain technology is set to be one of the key drivers behind the disruption of the technological landscape in the near future~\cite{swan2015blockchain, underwood2016blockchain}. Decentralized technologies are also the cornerstone behind the Internet 3.0 and Industry 4.0 revolutions that are undergoing~\cite{lu2017industry}.

Blockchain technology is already a driver behind decentralized and distributed IoT systems, providing security~\cite{qian2018towards}, trust~\cite{song2018blockchain, hammi2018bubbles}, data management~\cite{ayoade2018decentralized}, peer-to-peer transactions~\cite{chen2018devify},  and fault-tolerand middlewares~\cite{su2014decentralized}. Blockchain platforms can be divided in two main types depending on how they manage user credentials, which have a direct impact on their applicability: (i) permissionless, or public, and (ii) permissioned, private, or consortium, blockchains. They differentiate in that public blockchains are based on anonymous nodes with equivalent status, while consortium or private blockchains introduce different types of nodes and permissions, some of which require authentication in order perform certain actions. While trust in permissionless blockchains is shared and distributed, in permissioned blockchains there is a series of validator nodes that represented trusted authorities~\cite{zheng2017overview}.

One of the main issues stopping a wider adoption of blockchain in IoT systems is scalability, an inherent problem to Bitcoin's architecture that multiple researchers have been addressing~\cite{vukolic2015quest, karame2016security}. While smart contracts have great potential in the IoT and distributed systems in general, their scalability and performance is closely tied to the overall performance of blockchain systems~\cite{scherer2017performance}. Nonetheless, multiple advances in recent years have demonstrated that novel technologies can bring significantly higher degrees of scalability and performance to next-generation blockchain systems. Among these, Elastico provided the first implementation of a sharding protocol in  a permissionless blockchain~\cite{luu2016secure}. Sharding is a technique that enables the distribution of nodes in a blockchain into subchains for performing parallel validation, thus increasing throughtput and reducing latency. A more recent scalable blockchain is OmniLedger~\cite{kokoris2018omniledger}, which reports better scalability than Elastico and promises VISA-level latency and throughout if enough nodes form up the network. 

Owing to the distributed nature of blockchain systems, and distributed ledger technology (DLT) in general, IoT systems integrating them must already have a distributed architecture by themselves. Therefore, it is only natural that blockchain is integrated at the edge layer in most occasions, which represents the most distributed and interconnected layer of a typical IoT system. While sensors and actuators could be considered more distributed, they are not necessarily capable of node-to-node communication. Through this chapter, we utilize the terms blockchain and distributed ledger equivalently. However, distributed ledger technology (DLT) is often utilized to include more general systems that do not implement blockchains \textit{per se}, but instead rely on some other type of network or data management architecture. An example of this is IOTA, which utilizes acyclic directed graphs representing more general data structures. The rest of this introduction delves into more details behind the nature of mobile edge computing and its integration with blockchain/DLT technology.

\subsection{MEC and Network Slicing}

The European Telecommunications Standards Institute (ETSI) has promoted the standardization of Multi-Access Edge Computing (MEC)~\cite{etsi2018mec}, which shares the acronym with Mobile Edge Computing (MEC). The "multi-access" term puts an emphasis on the multi-tenant infrastructure and better reflects non-cellular operators~\cite{shahzadi2017multi, etsi2018mec}. In this chapter, we do not make distinctions between the two terms as our focus lays on the role of blockchain with edge computing. MEC standardization has been led by the MEC Industry Specification Group (ISG) since the end of 2014. One of the main objectives of the ETSI MEC ISG is to define the base technologies for distributed and multi-tenant clouds that are meant to be deployed at the edge of the radio access network (RAN)~\cite{hu2015mobile}. By deploying data aggregation and processing tasks directly at the edge of the network, MEC services can provide better reliability, lower latency and higher-throughput~\cite{taleb2017multi, qingqing2019odometry, queralta2019edgeai}. We will specifically discuss throughout this paper how blockchain technology can play a key role in terms of security and robustness for the resource management needed in a multi-tenant edge infrastructure, as well as enhance the services that MEC applications can provide~\cite{zhu2018edgechain, xiong2018mobile, queralta2020blockchain, rahman2018blockchain}.

One of the key architectural cornerstones enabling multi-tenancy and co-existing verticals at the MEC layer is network slicing~\cite{nextgen2016study}. Network slicing provides the base for interfacing blockchain with other MEC services for a wide array of application scenarios~\cite{queralta2020enhancing}. Network slicing refers to the co-existence of multiple software defined systems and networks (slices) sharing a common hardware infrastructure. Each of the slices can be thus designed independently and optimized for a particular application or business vertical~\cite{alliance2016description}. In particular, slicing for vehicular communication and offloading, together with 5G-and-beyond connectivity, are set to define the mobility of the future~\cite{giust2018multi}.

\subsection{Integration of Blockchain and MEC}

The integration of blockchain within the MEC layer has been object of extensive research over the past few years. Systems integrating blockchain and edge computing can be roughly divided among those in which edge services are part of a larger blockchain system~\cite{casado2018blockchain, nawaz2019icmu, liu2019joint}, and those in which blockchain is one of the services enhancing edge services~\cite{zhu2018edgechain, queralta2020endtoend, rahman2018blockchain, dai2019blockchain,  luong2018optimal, queralta2020blockchain}. In this chapter, we are particularly interested in the latter type, as blockchain can provide a key piece in enabling truly distributed, secure and efficient edge computing. With monetization of MEC being a central topic of discussion since its early proposal~\cite{taleb2017multi}, multiple works have focused towards either enhancing security or utilizing blockchain as a marketplace framework for users to access different applications at the edge~\cite{xiong2018mobile, rahman2018blockchain, dai2019blockchain}. More recently, other works have also delved into the potential of blockchain as a framework for managing edge resources~\cite{samaniego2016hosting, samaniego2016using, samaniego2017virtual, luong2018optimal}, as well as supporting autonomy in distributed robotic systems~\cite{queralta2020enhancing}.

From the security point of view, the integration of blockchain technology brings evident benefits to edge computing. Among the main threats identified in a recent report from the European Union Agency for Cybersecurity (ENISA) on 5G networks and edge infrastructure~\cite{enisa}, blockchain and DLT technologies can help address multiple remaining challenges. For instance, permissioned DLTs with built-in identity management naturally provide an extra layer of resilience against malicious diversion of network traffic, manipulation of traffic, or authentication traffic spikes. When blockchain tehcnology is applied to resource management, it can serve as a framework to mitigate risks in terms of abuse of third party hosted network functions, manipulation of the network resources orchestrator, or opportunistic and fraudulent usages of shared resources, among others. Moreover, safety-critical applications can benefit from the enhanced security that blockchains and other DLTs provide. These include the automotive sector with vehicle to everything communication routed at the edge~\cite{liu2018blockchain, kang2018blockchain}, and the healthcare sector~\cite{rahman2018blockchain, gia2019blockchain, ferrer2018robochain}.

\subsection{Chapter Structure}

Multiple surveys and review papers have recently been published on the convergence of blockchain and mobile edge computing~\cite{yang2019integrated, danzi2019delay, khezr2019blockchain, nguyen2020blockchain}. Other surveys in either the blockchain or edge computing domains also mention the potential for integrating one with another~\cite{gao2018survey, joshi2018survey, khan2019edge, moura2020fog}. In these and other works, scalability is often identified as one of the key aspects limiting the adoption of blockchain in edge computing. Nonetheless, these works describe the scalability problem either as a systemic blockchain problem~\cite{yang2019integrated}, or from a system point of view~\cite{khezr2019blockchain}. Most works also focus on a specific blockchain, Ethereum being the most widely researched blockchain for IoT~\cite{danzi2019delay}. In a blockchain, consensus algorithms are the main bottleneck in terms of scalability, i.e., the mechanisms enabling all nodes in the blockchain network to validate transaction and stay synced. Depending on the type of consensus algorithm, the scalability of the system might be limited by either the computational complexity of the algorithm, or its communication complexity. We believe there is a gap in the literature describing how the consensus algorithms affect the scalability from these two points of view. Our objective is to bring further insight in this area, providing a literature review and a discussion on the topic.

In this chapter, we introduce the main consensus algorithms that form the backbone of different blockchain solutions, including newer generation distributed ledgers that do not follow many of the paradigms defined within the Bitcoin and successive blockchains. We then describe what can be the role of edge computing when it integrates blockchain/DLT systems. In particular, we discuss the potential for the different solutions in the IoT, from the point of view of scalability but also discussing the different applications that are most suitable for different blockchain/DLT solutions. We do this from the point of view of consensus algorithms and their computational and communication complexity. Compared to previous works surveying the integration of blockchain and edge computing~\cite{yang2019integrated}, we provide a novel classification of current research directions from an architectural point of view (Section 3), while giving more insight into how the different consensus algorithms affect the integration of blockchain/DLT and edge computing (Section 4).

The rest of this chapter is organized as follows. In Section 2, we introduce the main consensus algorithms in blockchain systems and other DLTs, together with the most prominent results in highly-scalable and low-latency blockchains. Section 3 then reviews specific applications of blockchain at the MEC layer, and discusses how the different consensus protocols integrate at the edge. In Section 4, we discuss on the best blockchain/DLT solutions for different applications in the IoT, and how next-generation systems that are currently under development might change the IoT and MEC landscape. Finally, Section 5 concludes this work.

\section{Blockchain Technology: an Evolving Paradigm}

In this section we start with the basics of blockchain technology and move into how the field is evolving towards lower-latency, higher-throughput, and new concepts aimed at increasing flexibility and scalability, such as sharding. We provide a historical point of view on the different consensus algorithms that have been proposed for blockchains and other distributed ledgers, and include an overview of the most prominent so-called third-generation blockchains.

\begin{figure}
    \centering
    \includegraphics[width=\textwidth]{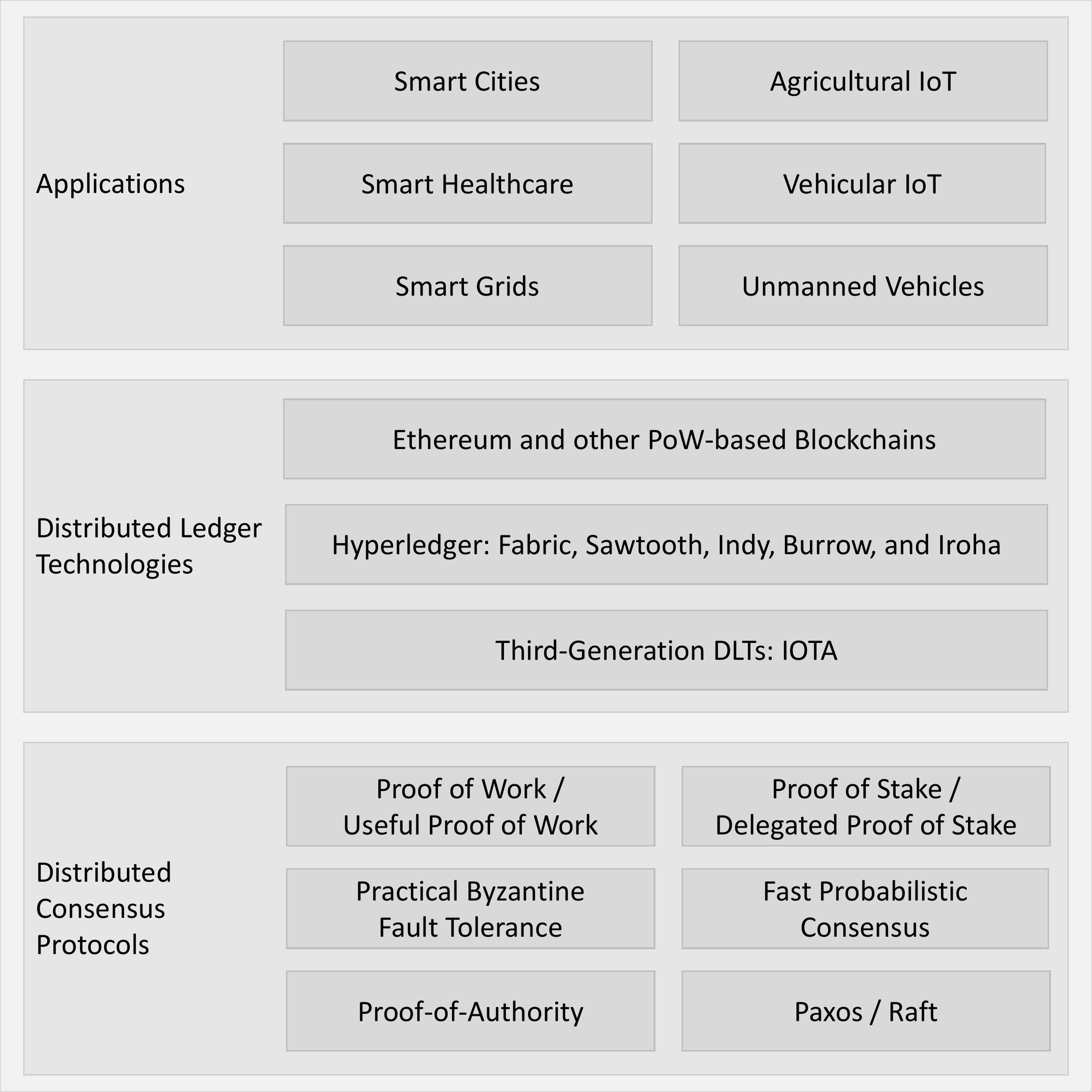}
    \caption{Blockchain/DLT Consensus protocols, systems, and applications in integration with the Internet of Things.}
    \label{fig:layers}
\end{figure}

Consensus mechanisms are one of the key aspects within the design of decentralized networked systems or distributed computing systems. Consensus mechanisms are those algorithms that enable multiple independent agents to reach an agreement on a certain value, operation, transaction, or other types of data. In a distributed and decentralized system, different agents, or nodes, need to be able to trust each other. Consensus mechanisms are the enablers of trust among agents. The most popular consensus mechanisms to date in blockchain systems, according to a survey from Li \textit{et al.}~\cite{li2017survey}, are proof of work (PoW), proof of stake (PoS) practical byzantine fault tolerance (PBFT) and delegated proof of stake (DPoS), with other significant approaches including proof of authority (PoA), proof of elapsed time (PoET) or proof of bandwidth (PoB). Apart from some of the more traditional consensus algorithms listed above (e.g. PoW utilized in Bitcoin or Ethereum, and PoS being part of Ethereum 2.0 plans), in this document we also review consensus protocols utilized in third- and fourth-generation distributed ledger systems such as the fast probabilistic consensus (FPC), and the cellular consensus (CC). We also put an emphasis on defining the key technologies behind IOTA, a DLT designed for the IoT and an ideal candidate for integrating DLTs with edge computing.

\subsection{Proof of Work}

Nakamoto's proof of work designed for Bitcoin~\cite{nakamoto2008bitcoin} has heavily influenced the development of new solutions for newer-generation blockchain systems. The PoW implementation in Bitcoin was a new application for an old algorithm. Originally proposed by~\cite{dwork1992pricing} as a solution to deter spam activity from email senders, the main idea behind PoW systems has remained unchanged: to request to all networked agents to solve computationally intensive cryptographic problems in order to validate their activity, their identity, or those of another agent. In general terms, a PoW algorithm is, at its most fundamental level, an algorithm that solves a cryptographic problem with a solution that is, in relative terms, hard to find and easy to validate. The computational complexity of the validation of a PoW solution is therefore considerably smaller than the complexity of finding such solution.

Ethereum, the second most popular blockchain system after Bitcoin, also relies on PoW-based consensus to validate new blocks in the blockchain. A block can be roughly defined as each of the entries in the distributed ledger that blockchains implement. A block does not include a single transaction, but often a set of transactions that are near in time. These transactions represent the block's body, where transactions are defined in a generic manner and do not represent only the exchange of cryptocurrencies. Transactions in PoW-based blockchains are not validated individually, but instead all the transactions in a block get validated when the block containing them is validated itself. A block is validated, or mined, by solving a PoW puzzle. The original and most widely used puzzle in blockchains can be summarized as follows: the PoW algorithm must find a block header, which is the result of applying a cryptographic hash function to the content of the block body, satisfying some predefined condition. However, for a fixed hash function and a fixed block body, the resulting hash will always be the same. In order to meet this condition (e.g., finding a hash smaller than a certain value), the algorithm must then find some other value, called a nonce, to be added to the current block body. Finding a nonce is the process often called mining. Once a block is mined, it is added to the blockchain and all other agents in the network can validate the solution. In Bitcoin and other blockchain systems, the miner of a block gets a reward in the form of new cryptocurrency, thus motivating nodes to participate in the transaction validation process.

One of the problems of PoW-based blockchains is that two agents could solve a PoW puzzle at virtually the same time, for the same or different nonces. This can create two branches, or forks, in the blockchain. Nodes are situated in the branch of the solution that they received first. In Bitcoin, a built-in policy establishes that if one fork is longer than the other (or it accumulates more cryptographic complexity), then all agents in the network judge it as the authentic one. This is a practical solution as it is highly improbable that two consecutive blocks will be solved simultaneously by two pairs of nodes. In any case, even if two or more blocks are solved at the same time, at some point one of the forks will become longer. This defines the so-called 51\% or double spending attack, as malicious nodes would need to to control at least 51\% of the network's computing power in order to be able to introduce a faulty transaction in a block, validate it, and keep validating consecutive nodes in the corresponding fork so that it is accepted as the canonical fork by the network. When the size of the network and the number of miners increases, the probability of such attack is reduced, thus giving the blockchain its immutability and data integrity properties. 

The benefit of having an expensive PoW solution in terms of hardware, energy consumption and time is that it is equally expensive for malicious nodes to attack the network. Part of the security of PoW thus comes from disincentivizing attackers because of the large a priori investment required in order to be able to attack and gain control of the network, which would not pay off even if the attack is successful~\cite{nguyen2018survey}. 

\subsection{Proof of Useful Work}

Part of the research community has argued that taking into account the humongous amount of computational resources and electric energy put into mining to solve PoW puzzles, at least these could be defined in a way that the solutions found would help research in other fields. As an example, King \textit{et al.} proposed the definition of PoW puzzles that would find long chains of primes~\cite{king2013primecoin}. Solving these PoW would be then dedicated to solve a mathematical problem which consists on finding the distribution of the Cunningham prime chain. In this case, the Fermat Primality Test would be used to validate the PoW solutions.

A different research approach is the definition of simpler PoW requiring less computational resources in order to reduce the entry barrier and provide a more uniform distribution of mined currency. Pagh \textit{et al.} introduced the concept of Cuckoo hashing, in which the PoW difficulty would remain constant over time~\cite{pagh2004cuckoo}.

\subsection{Proof of Stake}

The basis for security and robustness in a PoW system comes from the amount of computational resources needed in order to gain control over the network. Nonetheless, this computational complexity also brings limitations. First, it limits the probability for news nodes to be able to mine new cryptocurrency by themselves if they join a large network. Second, it also limits the number of transactions that can be validated within a certain time interval. For instance, in Bitcoin, it takes an average time of 10 minutes to validate a block and all the transactions it includes~\cite{barber2012bitter}. A different consensus approach that does not rely on computational complexity and that has gained momentum in recent years is Proof of Stake (PoS). One of the main objective of PoS systems, which is being introduced, for instance, as part of Ethereum 2.0, is to reduce transaction validation latency. One of the first implementations of PoS in a blockchain system, which showed clear benefits in this direction, was demonstrated with Nxtcoin~\cite{popov2016probabilistic, nxt2018whitepaper}. The idea behind PoS is to value the cryptocurrency that validating nodes put \textit{at stake}, instead of their computational power. PoS mechanisms elect validators with a probability proportional to the size of their stake, which is often closely related to the amount of cryptocurrency that the node, or miner, owns. Nodes can lose the total value of their stake if they incur in fraudulent validations. In~\cite{bentov2016cryptocurrencies}, a similar PoS system was proposed where the probability of selection of the nodes validating transactions was calculated based on both the pure stake and the state of the block being validated in the blockchain.

The 51\% attack discussed in the PoW consensus mechanism is still a potential attack vector in a PoS system. However, while in the PoW case attackers need to obtain control over 51\% of the network's computing power, which becomes increasingly easy as larger pools monopolizing the mining process are created, in a PoS system an attacker needs control over 51\% of the cryptocurrency's total supply. This is, in theory, a more difficult problem than gathering enough computing power. 

Owing to the significant reduction of the computational complexity of the consensus algorithms with PoS when compared to PoW, the energy consumption footprint is also reduced. PoS thus provides a more energy-friendly alternative which in turn enables nodes with lower computational capabilities to participate in the blockchain as equals to all others. Multiple authors, such as \cite{o2014bitcoin} or \cite{de2018bitcoin}, have studied the sustainability of Bitcoin's growth and its energy footprint, which researchers estimate to be the equivalent, on a yearly basis, to non-renewable energy resources consumed by entire nations of the size of Czech Republic or Jordan. Nevertheless, this also means that because miners do not need to dedicate large amounts of computational resources to mining, it is easier to perform Sybil attacks spawning multiple identities within a single malicious node.

In general terms, a PoS system relies on a validator or a set of validators which are eligible after depositing part of their stake. In other words, as described by Buterin \textit{et al.}~\cite{buterin2019incentives}, nodes earn the right to propose a block only after locking part of the coins they own on the blockchain. This is an extended definition over the pure PoS system firstly implemented in~\cite{king2012ppcoin} as part of PPCoin, in which the total miner's stake is directly considered.

\subsection{Practical Byzantine Fault Tolerance}

The Practical Byzantine Fault Tolerance (PBFT) consensus algorithm was first proposed by Castro \textit{et al.} in 1999~\cite{castro1999practical}. PBFT was the first algorithm with the ability to operate in large asynchronous networks such as the Internet, while providing over one order of magnitude in processing power improvement over previous methods, allowing for high-performance Byzantine state machine replication, and demonstrating thousands of requests per second. Byzantine fault tolerance can be described as the capacity of a system to maintain proper operation when multiple errors or unexpected behaviour occur within part of the system, but not its totality~\cite{keichafer1988maft}. In a distributed network and considering the consensus problem, this is equivalent to the ability of the network to provide a robust consensus even in an scenario where a subset of nodes act maliciously, failing to forward valid data or sending invalid information. 

In a PBFT system, nodes are distinguished between validating and not-validating peers~\cite{sousa2018byzantine}. The validating nodes run the consensus algorithm, in which they replicate a state machine and evaluate its result. A client makes a request that is transmitted over the peer-to-peer network through the non-validating nodes, which act as proxies between clients and validators. Non-validating nodes do not participate in the consensus mechanism, but are able to confirm the results. The PBFT algorithm is able to provide consensus across the network when at most one third of the nodes behave arbitrarily or maliciously. Because the validator nodes need to arrive to the same results regarding the client request, the state machine that is replicated must be deterministic. 

In comparison with PoW and PoS systems, in PBFT individual transactions can be confirmed without the need to wait for a block including several transactions to be added to the blockchain. In terms of energy efficiency, PBFT requires less computational resources than a PoW consensus, but increases the probability of a Sybil attack, where a malicious node would create multiple instances pretending to be a large number of parties. In practice, PBFT is often combined with a PoW that must be solved in order to join the network and within certain time intervals to ensure that every node in the network is dedicating some minimum computational resources to the collective validation effort. An important benefit of PBFT over PoW and PoS is the low reward variance, as every node can be incentivized. This lowers the reward variance across miners. Nonetheless, the scalability of PBFT is an issue due to the large number of peer-to-peer communication exchanges required.

\subsection{Third-Generation DLTs - Beyond Blockchain}

Excluding Bitcoin and Ethereum, which represent the majority of the cryptocurrency market capitalization, one of the most successful blockchains within the IoT and industrial domains has been Hyperledger~\cite{cachin2016architecture}. Launched in 2016 by the Linux Foundation, the Hyperledger project is divided in five main subprojects where blockchain frameworks for different aims are being developed: Fabric, Sawtooth, Indy, Burrow, and Iroha~\cite{saraf2018blockchain}. Among these, Hyperledger Fabric is the most popular, an enterprise-level and production-ready permissioned distributed ledger framework that has already been applied across various industrial fields~\cite{androulaki2018hyperledger}. The aims behind the project include open-source and cross-industry development of an scalable framework for smart contracts. Through the rest of this chapter, we utilize Hyperledger to refer to Hyperledger Fabric unless otherwise specified.

The consensus mechanism utilized in Hyperledger vary depending on the subproject. For instance, Hyperledger Fabric relies on RAFT~\cite{ongaro2014search}, while Hyperledger Indy utilizes Plenum, based on Redundant Byzantine Fault Tolerance (RBFT)~\cite{aublin2013rbft}. Different blockchains following the hyperledger design ideas rely on PBFT or adapted BFT approaches.

In recent years, blockchain technology has evolved towards a wider range of network definitions that do not keep the original structure of a blockchain in terms of how to store data within a distributed ledger. Among these, one of the most prominent distributed open ledgers under development is IOTA~\cite{popov2018tangle}. IOTA's backbone is a directed acyclic graph that defines the \textit{tangle}. The tangle is the underlying network upon which IOTA is built. While Bitcoin was born mainly as a distributed cryptocurrency, Ethereum evolved from it into a platform for smart contracts, and Hyperledger is intended for industrial use, IOTA was specifically designed with the IoT in mind~\cite{divya2018iota}. In IOTA, there are no miner or validator nodes confirming transactions, but instead each user must participate in the validation of two transactions before being able to issue a new one on its own. This approach, together with the tangle's structure, makes IOTA highly scalable and free to use. IOTA's development is open-source and led by the IOTA foundation.

IOTA's consensus protocol is defined within the Concordice system~\cite{popov2020coordicide}. The main differentiating aspect of IOTA's tangle is the fact that multiple disconnected subnetworks can coexist for certain periods of time. This means, for instance, that while a blockchain cannot contain two conflicting transactions in committed blocks, the tangle might temporarily contain two such transactions. IOTA deals with this, however, in a similar manner as Bitcoin does: the fact that a transaction is included in the blockchain does not automatically mean it is valid, as two forks of the chain might exist until one is deemed longer and this valid. Therefore, in both cases there is only information about the \textit{probability} of a transaction being valid, which increases as the blockchain, or the tangle, grow after that given transaction. In order to make a decision on two conflicting transactions in IOTA and reach a consensus across the network, Concordice proposes two consensus protocols: the fast probabilistic consensus (FPC) and the cellular consensus (CC). FPC, introduced in~\cite{popov2019fpc}, is a leaderless probabilistic binary consensus protocol. FPC has low complexity from the communication point pf view, and is robust in a Byzantine infrastructure. As with PBFT, the basic idea behind FPC is voting. In any case, IOTA is still under development and is not production-ready. More detailed information on IOTA's consensus and CC is available in~\cite{ramosreview} and~\cite{nelson2020majority}.

Other DLT solutions claiming to be third-generation blockchain are Nano~\cite{lemahieu2018nano}, with its underlying block lattice, and Skycoin~\cite{skycoin}, aimed at powering the Web 3.0. While Nano and IOTA are recent technologies, Skycoin has been under development for several years and was born out of a series of external audits into Bitcoin, which revealed the different flaws in the PoW consensus protocol.

\subsection{Smart Contracts}

Second-generation blockchain systems, largely represented by the Ethereum blockchain, were defined as those introducing the ability of executing distributed programs within the blockchain itself, therefore extending their applicability beyond cryptocurrency transactions and into the validation of more general types of transactions. These programs that can be executed within a blockchain are called \textit{smart contracts}, with one of the most notorious implementations being part of the Ethereum Virtual Machine and its corresponding stack~\cite{wood2014ethereum}, which provides a Turing complete language as part of its framework~\cite{hildenbrandt2018kevm}. Ethereum also introduced a new programming language to be dedicated to the development and implementation of smart contracts: Solidity~\cite{solidity}. Smart contracts as defined with Solidity code can be seen as a set of instructions defining transitions between states of the program, with both the data representing the different states and the code defining the transitions being stored at specific addresses within the Ethereum blockchain.

In Ethereum, smart contracts are part of the Ethereum Virtual Machine (EVM)~\cite{dannen2017introducing}. 
The EVM is based on the existence of contract accounts in the blockchain, which extend the functionality of external accounts, those controlled by a human or network node through a public-private key pair. Contract accounts operate in an automated way as a function of the code stored within the account. While external accounts are defined based on their key pair, with an address determined based on the public key being assigned to each node joining the network, contract accounts have addresses that are determined when the contract is created. In Ethereum, the address space is shared among both types of accounts. Contract accounts are created through transactions that have a null or empty recipient. Those transactions must contain code that outputs the smart contract's code, which is then generated when the transaction's code is executed within the EVM. In general terms, transactions including a payload and Ether (Ethereum's cryptocurrency) between external accounts in Ethereum are extended so that when a transaction's target account is a contract account containing a set of code instructions, these are executed given the payload in the transaction. A key concept in Ethereum is gas. Upon creation, transactions are assigned a definite quantity of gas. The gas is a measure of the processing power that will be dedicated to that transaction. In other words, the gas is the transaction fee. The gas is initially charged into the transaction, and its reserve gradually decreases as a function of a set of predefined rules when the EVM executes the different transaction instructions. The gas that is left is refunded to the transaction creator. The gas price, which is paid upfront, is decided by the creator node. Miners, which obtain the gas price as a reward, decide which transactions to mine based on the amount of gas included. Therefore, the gas price is decided based on the market and the desired priority for a specific transaction. 

\subsection{Sharding and Scalability}

While second-generation blockchains introduced new functionality and improvements over Bitcoin-based blockchains at different levels, one of the main challenges in blockchain systems remained: scalability~\cite{vujivcic2018blockchain}. This is mostly due to the large and increasing amount of computational resources required for mining. From the communication point of view, Bitcoin and other similar blockchains only require one broadcast per block, and therefore the main bottleneck comes from computation (which cannot be directly decreased while maintaining security). In PBFT-based systems, multicast messages are required for validation, and thus the main scalability problem is the communication cost~\cite{vukolic2015quest} (which cannot be directly reduced either without compromising security and robustness of the consensus mechanism). Multiple research efforts have been directed towards the realization of more scalable systems, with new blockchains based on PoS and PBFT showing promising results. Elastico, introduced in~\cite{luu2016secure}, was one of the first scalable blockchains that introduced the concept of sharding: to divide the network in subnetworks, or \textit{shards}, that would validate transactions in parallel. Elastico was the first blockchain system to provide a full implementation of a sharding scheme for a permisionless blockchain. A different early sharding proposal was presented in~\cite{merklix}, where Merklix trees are utilized to merge the state of the different shards into the global blockchain state~\cite{introducingmerklix, qin2017cecoin}. 

Another blockchain system aimed at scalability that has had an important impact on subsequent research is OmniLedger~\cite{kokoris2018omniledger}. Omniledger scales linearly with the number of nodes in the blockchain, and reports transaction times able to match credit card standards if the size of the network arrives to a certain threshold. The key difference with Elastico in terms of scalability is that in Elastico the network performance scales with the computational power in a linear fashion, while in Omniledger it does so with the number of validator nodes. In Hyperledger, the scalability of the network has seen significant improvements since the release of Fabric 1.1.0~\cite{ferris2019does}. Moreover, the number of channels can be scaled with little to no impact on performance according to the same report.

Perhaps the biggest effort that is currently being put into the development of a truly decentralized, permissionless and scalable yet secure blockchain is the design and development of Ethereum 2.0~\cite{buterin2014next}, where huge amounts of computing resources will be no longer required for mining~\cite{ethereum20specification}. The Ethereum Foundation and other developers behind Ethereum 2.0 have embraced Proof of Stake as the main consensus mechanism, while still utilizing PoW to secure the network, and the concept of sharding towards scalability. The consensus is based on the Casper protocol~\cite{buterin2017casper}, which incentives for mining have been described in~\cite{buterin2019incentives}. The impact that shards have on transaction scalability is relatively clear, with a much larger throughput being possible in terms of transactions validated per second. Nonetheless, it is not straightforward to extend the implementation of smart contracts with sharding. As smart contracts have associated a series of data states corresponding to their code, each state change can be though of as a transaction. Contracts can be executed within a single shard, or a cross-shard synchronization mechanism must exist to allow for data to flow between shards. In~\cite{kokoris2018omniledger}, the authors introduced introduced Atomix, a client-driven lock/unlock protocol, to ensure that a single transaction can be committed across multiple shards, while enabling the possibility of unlocking rejected transaction proofs in specific shards. The original Atomix state machine can be extended to accommodate the execution of smart contracts across shards. 

\section{Blockchain Technology for Mobile Edge Computing}

\begin{figure}
    \centering
    \includegraphics[width=\textwidth]{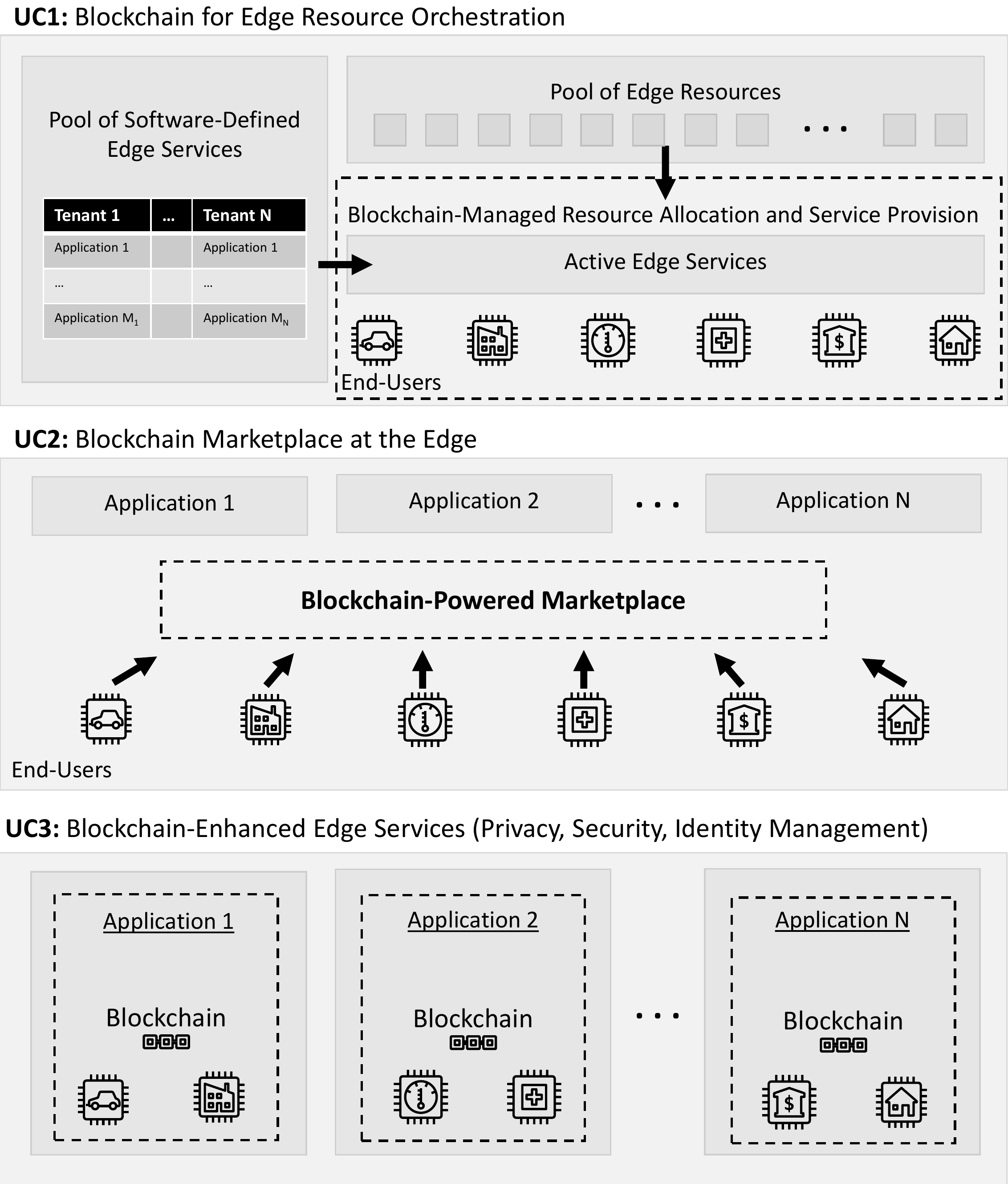}
    \caption{Main use cases for blockchain within edge computing systems. (UC1) Blockchain-powered resource allocation and service provision; (UC2) Blockchain-powered marketplace for interfacing users and services; (UC3) Blockchain-enhanced individual edge services relying on blockchain technology for security, privacy, data management and audits or identity management, among others.}
    \label{fig:use_cases}
\end{figure}

This section reviews and classifies the existing research in the integration of blockchain and MEC from an architectural point of view. We classify the different approaches on three main categories, illustrated in Fig.~\ref{fig:use_cases}. The first category encompasses works providing a system-level integration where a blockchain is one of the key pieces at the heart of the edge infrastructure, managing services and resources. The second category includes approaches that utilize blockchain as a middleware between the edge infrastructure (hardware and software) and the third-party services being provided through MEC. Finally, the last category comprises those works where the blockchain is part of individual applications, for aspects such as security or identity management.

In general terms, Ethereum is the most widely applied blockchain platform in the IoT, owing to the maturity of its smart contracts framework enabling complex interactions between data producers and consumers~\cite{huh2017managing, pustivsek2018approaches}. In the same area, Hyperledger has potential to disrupt the IoT with more scalable solutions and the ability to run distributed programs as chaincode~\cite{valenta2017comparison}. In all these cases, nonetheless, the blockchain runs in embedded edge gateways providing stable connectivity, and where enough power and computational resources is available. With the potential to reach embedded devices at the sensor layer, and being developed specifically for the IoT, IOTA is set to play an increasingly important role. Owing to its low inherent computational requirements and being highly scalable, IOTA is the ideal candidate for edge computing systems and hardware. 

\subsection{MEC Resource and Service Orchestration with Blockchain}

One of the most critical points at the edge is resource orchestration~\cite{taleb2017multi}. In order to enable a wide variety of use cases, multi-tenant applications, and ad-hoc deployment of different modules, MEC infrastructure needs to be able to manage its resources in real time, while also orchestrating how the network is being utilized. This includes processes from allocating hardware resources for the different virtualized applications to managing the spectrum or the bandwidth that might be in use for computational offloading by different service providers.

Blockchain technology can provide multiple advantages to orchestration at the edge: enhanced security and identity management, together with distributed consensus algorithms to implement the resource allocation decision processes. In this area, EdgeChain was introduced by Zhu \textit{et al.} as a middleware platform to deploy third-party applications across the MEC layer~\cite{zhu2018edgechain}. In~\cite{queralta2020blockchain}, the authors introduce a blockchain framework that relies on smart contracts for managing network bandwidth and resource allocation in a distributed and collaborative computational offloading framework. In~\cite{queralta2020enhancing}, a similar idea is extended towards managing network infrastructure and the available computational resources focused at enhancing autonomy of self-driving cars and other autonomous robots forming distributed robotic systems. In this paper, the blockchain MEC slice was the key slice managing the deployment of applications across other MEC slices supporting different verticals within the automotive sector. Further adoption of blockchain for computational offloading will require, however, higher-bandwidth and lower-latency blockchain frameworks enabling real-time sensor data to be streamed for applications such as autonomous mobile robots~\cite{qingqing2019odometry, qingqing2019monocular}.

Resource orchestration processes have underlying optimization algorithms that can be implemented either in a more traditional deterministic manner, or relying on machine learning models. Several authors have proposed the utilization of deep reinforcement learning for computational offloading in blockchain-powered edge computing. In~\cite{qiu2019online}, the authors demonstrate an approach that is able to improve long-term performance in a computational offloading scheme, with an adaptive genetic algorithm to improve the exploration processes while learning. In~\cite{dai2019blockchain}, the authors describe different situations in which blockchain can support resource management at the edge with deep reinforcement learning: spectrum sharing, vehicle-to-vehicle energy trading, computational offloading, or device-to-device content caching.

\subsection{Blockchain for a MEC Services Marketplace}

DLT can also provide a platform for building a marketplace between end-users and third-party edge application through either a transparent, secure and auditable monetization framework or as a middleware for sharing data securely between producers and consumers. In the former direction, Xiong \textit{et al.} deployed a blockchain at the edge to enable resource-constrained devices producing data to sell it to third-party applications~\cite{xiong2018mobile}. The pricing scheme introduced in the paper models the interactions within the IoT as market activities and the blockchain represented the framework for regulation of such activities. Distributed marketplaces based on blockchain for MEC services often utilize Ethereum as a base and the InterPlanetary File System (IPFS) for data storage. Examples are available in~\cite{ozyilmaz2018idmob} or~\cite{ranganthan2018decentralized}. A study describing the different challenges and opportunities is available in~\cite{varghese2018realizing}.

\subsection{Blockchain-Powered MEC Services}

In~\cite{liu2018blockchain}, the authors describe how blockchain can play a key enabled role in interconnected vehicles from the security point of view. In particular, blockchain is exploited for data management, but also for energy management in electric vehicles, with the authors proposing blockchain inspired data coins and energy coins. An edge computing security scheme is proposed including these two interaction aspects. An approach more related to the nature of blockchain as a cryptocurrency framework was proposed by Liu \textit{et al.} in~\cite{liu2019joint}, where the authors present an offloading framework not for data but for the blockchain itself and related mining operations. In general, blockchain can support edge services by providing enhanced privacy and security~\cite{nawaz2019icmu}, decentralized data management~\cite{ayoade2018decentralized}, or identity management~\cite{ren2019identity}.

\section{Performance and Scalability of DLTs at the Edge}

In this section we describe the benefits and drawbacks of the different consensus protocols and DLT solutions for each of the three main use cases defined in the previous section and illustrated in Fig.~\ref{fig:use_cases}, as well as for edge computing in the industrial internet of things. A basic classification of some of the protocols introduced in the previous section from the point of view of the capabilities of embedded IoT systems is given in Table~\ref{tab:comparison}.

\begin{table}
    \caption{Comparison of consensus protocols in terms of their applicability within resource-constrained devices in the IoT.}
    \begin{center}
        \begin{tabular}{@{}lcccc@{}}
            \toprule
             & \hspace{1ex}\textbf{PoW}\hspace{1ex} & \hspace{1ex}\textbf{PoS/DPoS}\hspace{1ex} & \hspace{1ex}\textbf{PBFT}\hspace{1ex} & \hspace{1ex}\textbf{Concordice}\hspace{1ex} \\
            \midrule
            \textbf{Computationally-Constrained Devices}\hspace{2em} & \xmark & \cmark & \xmark & \cmark \\
            \textbf{Communication-Constrained Devices} & - & \xmark & \xmark & \cmark \\
            \textbf{Intermittent Connectivity} & \xmark & \xmark & \xmark & \cmark \\
            \textbf{Independent Subnetworks} & \xmark & \xmark* & \cmark** & \cmark*** \\
            \textbf{Production-Ready Platform} & \cmark & \cmark & \cmark & \xmark \\
            \bottomrule
        \end{tabular}
    \end{center}
    \footnotesize{*Recent proposals implementing sharding might be considered subnetworks, however here we refer to the ability of specifically creating a subnetwork from a given set of nodes. \\**Channels in Hyperledger enable data separation but need to remain connected to the main net. \\\***The tangle in IOTA enables sets of nodes to be disconnected for certain periods of time and rejoin the network later on.}
    \label{tab:comparison}
\end{table}

\subsection{Blockchain Technology in Resource-Constrained Devices}

Consensus protocols in the different DLTs are the key performance indicators, and they are directly related to the minimum capabilities that nodes in the network must meet. In PoW-based blockchains, including Bitcoin and Ethereum, resource-constrained devices in the IoT that are potentially battery powered do not have the ability to participate as full nodes in the network. In Ethereum, nonetheless, the blockchain has adapted to some extent towards embedded IoT devices. For instance, the Zerynth Ethereum library provides basic capability to embedded microcontrollers running MicroPython~\cite{zerynth}. It enables sensor nodes to create signed transactions and execute contract calls.

Hyperledger Fabric and IOTA, designed with scalability in mind, do not have such strong computational requirements. The consensus protocols at the hearth of Hyperledger, however, have high communication complexity and therefore require nodes to be able to communicate frequently and with low-latency. Hyperledger can therefore run in embedded IoT edge gateways with wired internet connection but its extendability to wireless and potentially battery-powered sensor nodes is limited. In this area, IOTA has a comparative advantage. In particular, STMicroelectronics has collaborated with the IOTA foundation in the development of X-CUBE-IOTA~\cite{iotaxcube}, a complete middleware that enables IoT sensor nodes based on STM32 microcontrollers to build IOTA applications and access the IOTA distributed ledger directly.

In terms of communication-constrained devices, low-power wide area networks (LPWANs) have emerged in recent years as a solution for extending the range of applications, with LoRa and LoRaWAN being the most prominent radio and network technologies~\cite{queralta2019edgeai, queralta2019lpwan}. Edge computing is a natural paradigm to be integrated with LPWAN networks owing to the low-bandwidth available and thus the need to preprocess large amounts of raw data~\cite{sarker2019lora, gia2019africon, gia2020lorawan}. However, the integration of blockchain into LPWAN networks is not direct~\cite{ozyilmaz2017work}. Current efforts deploy the blockchain either at the LPWAN gateways, which often have wired internet connection, or at the back-end servers~\cite{lin2017using, durand2018resilient}. More interesting use cases will be possible when the blockchain nodes can be interconnected via low-bandwidth and high-latency LPWAN links, which might be soon possible with IOTA and STM.

\subsection{Application Scenarios}

From the point of view of edge computing as a system encompassing multiple independent applications, the simplest use case is such in which blockchains are managed by each application independently. This allows for the same orchestration algorithms to remain in place, as well as co-existence of blockchain-based and other applications running at the edge. Depending on the nature of each of the applications, all of the DLT solutions presented in this chapter might be applied. For general IoT systems where data is gathered from sensor nodes and transactions between either the user or the sensors and the application back-end (which may or may not be deployed entirely at the edge) are relatively simple, then IOTA stands out by providing free transactions. This can be a key differentiating point in applications where data is routinely gathered and does not have specific value. Because IOTA's consensus is built in a way that all nodes need to take the validator role before being able to commit transactions, nodes do no need an additional incentive to validate and therefore there is no need for a transaction fee as with other blockchain platforms. If more complex transactions are required, with either real-time interaction between users or a user and sensor data being processed, then smart contracts might be needed. Ethereum is by far the most extended and used blockchain platform for smart contracts, and therefore it would be natural to rely on it. This will be an even better solution when Ethereum 2.0 is available. Nonetheless, relying on Ethereum or similar solutions involves an extra transaction cost, due to the need for mining new cryptocurrency to compensate nodes participating in the validation process. Alternatively, private Ethereum networks can be deployed and infrastructure managed by the application developers. This is specially important in PoW-based systems, but also in PoS systems as otherwise nodes would have no incentives on putting their stakes at risk.

When blockchain is utilized to power a marketplace of services at the edge, the cryptocurrency that blockchains build upon might play a more important role with the introduction of monetization. In this sense, monetization does not necessarily refer only to paying for services, but can also encompass the edge resources that services rely on~\cite{taleb2017multi}. Similar to the previous case, the choice of DLT framework has a significant dependence on the type of data management and processing that needs to be done. For simple applications in which services and end-users are predefined and communicate independently, then IOTA can provide a fast and scalable framework, while Hyperledger could be an alternative if there is enough infrastructure set to sustain the blockchain and validate transactions. These applications can cover a wide variety of scenarios: paying a highway toll, exchange of information for coordination between autonomous cars, track-and-trace in the logistics sector, or providing digital identity to citizens in a smart city. In all these cases, a common denominator is that the transfers of value, or information, are small and frequent in time, and therefore there is not enough incentive to utilize other blockchain platforms such as Ethereum where transactions involve a fee. Hyperledger, nonetheless, is only a viable option if either public or private infrastructure supports its use without an impact on the end-user. For more complex applications, both Hyperledger and Ethereum provide extensive support for smart contracts and execution of distributed applications.

The last of the use cases presented in the previous section, and involving the most complex system-level integration of DLT technology at the edge layer is resource allocation and service provision. In this case, different optimization algorithms in which the resource orchestrator relies need to be implemented on top of the blockchain for transparent management of resources. The processes involved in dynamic resource allocation and service provision are complex and therefore require blockchains able of running smart contracts. Ethereum provides a suitable platform from the functionality point of view, but lacks the ability to scale and the low control over latency would significantly affect the real-time allocation of resources. Moreover, the computational power needed to validate transactions would reduce the availability of edge resources. Until Ethereum 2.0 or a more scalable solution is available, Hyperledger has multiple competitive advantages in this area.

A different application scenario that has not been directly covered in the previous section is the industrial IoT. Industrial scenarios often differentiate in that they operate on private networks. Moreover, safety-critical applications require more control over the network parameters as well as over the data management itself. In these directions, Hyperledger Fabric stands out, with design decisions targeting industrial use cases since its inception. Not only does a permissioned Hyperledger blockchain provide a secure framework for management of identities and network control, but it is the ability to separate data across channels that can provide wider adoption in privacy-critical and safety-critical use cases.
\section{Conclusion and Future Work}\label{sec:conclusion}

We have reviewed the most important consensus protocols in traditional blockchains and novel distributed ledger technologies, together with the different applications and use cases resulting of the integration of blockchain and edge computing. In particular, we have described how the underlying consensus protocols affect the applicability of the different DLT systems for edge computing, with an emphasis on the current research trends in terms of scalability and performance. We have outlined the main benefits and drawbacks of Ethereum, Hyperledger and IOTA in four main use cases: (i) orchestration of edge resources and services, (ii) implementation of a marketplace of edge services, (iii) enhancing security, privacy or identity management of individual edge services, and (iv) providing a framework for data management in the industrial Internet of Things.

\begin{acknowledgement}
    This work was supported by the Academy of Finland's AutoSOS project with grant number 328755.
\end{acknowledgement}

\bibliographystyle{unsrt}
\bibliography{ref}

\end{document}